\documentclass[aps,prd,nofootinbib,showpacs,superscriptaddress,12pt]{revtex4}
\usepackage{txfonts}
\usepackage{graphicx}
\usepackage{dcolumn}
\usepackage{bm}
\usepackage{amssymb}
\usepackage{latexsym}
\begin{document}

\title{\bf Holographic Dark Energy Characterized by the Total Comoving Horizon and Insights to Cosmological Constant and Coincidence Problem}
\author{Zhuo-Peng~Huang}
\email[Electronic address: ]{zphuang@nudt.edu.cn}
\affiliation{Department of Physics, National University of Defense Technology, Hunan 410073, China}
\affiliation{State Key Laboratory of Theoretical Physics (SKLTP)\\
Kavli Institute for Theoretical Physics China (KITPC) \\
Institute of Theoretical Physics, Chinese Academy of Sciences, Beijing 100190, China}
\author{Yue-Liang~Wu}
\email[Electronic address: ]{ylwu@itp.ac.cn}
\affiliation{State Key Laboratory of Theoretical Physics (SKLTP)\\
Kavli Institute for Theoretical Physics China (KITPC) \\
Institute of Theoretical Physics, Chinese Academy of Sciences, Beijing 100190, China}

\date{\today}

\begin{abstract}
The observed acceleration of the present universe is shown to be well explained by the holographic dark energy characterized by the total comoving horizon of the
universe ($\eta$HDE). It is of interest to notice that the very large primordial part of the comoving horizon generated by the inflation of the early universe makes
the $\eta$HDE behave like a cosmological constant. As a consequence, both the fine-tuning problem and the coincidence problem can reasonably be understood with the
inflationary universe and holographical principle. We present a systematic analysis and obtain a consistent cosmological constraint on the $\eta$HDE model based on
the recent cosmological observations. It is found that the $\eta$HDE model gives the best-fit result $\Omega_{m0}=0.270$ ($\Omega_{de0}=0.730$) and the minimal
$\chi^2_{min}=542.915$ which is compatible with $\chi^2_{\Lambda {\rm CDM}}=542.919$ for the $\Lambda$CDM model.
\end{abstract}

\maketitle

%%-------------------------------------------------------------------------------------------------------------------------------------

\section{Introduction}

Recent cosmological observations \cite{Riess:1998cb, Perlmutter:1998np, Spergel:2003cb, Tegmark:2003ud} have indicated that the observable universe enters into an
epoch of accelerated expansion. Within the framework of the general relativity, the acceleration can be phenomenally attributed to the existence of a mysterious
negative pressure component named as dark energy. A number of models have been proposed to explain such an exotic energy component, which is referred to some recent
reviews \cite{Copeland:2006wr, Li:2011sd}. The simplest candidate is a positive cosmological constant. Although it remains well for fitting to the observations, a
cosmological constant, however, is plagued with the fine-tuning problem and the coincidence problem \cite{Weinberg:1988cp}.

In the spirit of the holographic principle \cite{'tHooft:1993gx, Susskind:1994vu}, it is proposed that \cite{Cohen:1998zx} in the effective field theory the
ultraviolet (UV) cutoff $\Lambda_{\rm uv}$ is related to the infrared (IR) cutoff $L$ due to the limit set by the formation of a black hole. It is generally given
in terms of the natural units as follows
 \begin{equation}
    L^3\Lambda_{\rm uv}^4 \lesssim LM_p^2  ~,
 \end{equation}
where  $M_P$ is the reduced Planck constant $(M_p^2=1/(8\pi G)$ with $G$ the Newton's~constant. This corresponds to the assumption that the effective theory
describes all states of the system, except those that have already collapsed to a black hole.  Due to such a dramatic depletion of quantum states, the vacuum energy
density caused by the quantum fluctuation is given by
 \begin{equation}
    \rho_{\rm vac} \sim \Lambda_{\rm uv}^4 \sim  M_p^2L^{-2}  ~. \label{vac}
 \end{equation}
It is interesting to note that if the IR cutoff $L$ is comparable to the current Hubble radius, the resulting quantum energy density $\rho_{\rm vac} \sim
\Lambda_{\rm uv}^4$ requires no magnificent cancelation with a bare cosmological constant in order to be consistent with observational bounds. Thus the fine-tuning
problem may reasonably be understood \cite{Cohen:1998zx} if such a correlation can consistently be realized, and it becomes natural to assume that such a vacuum
energy plays the role of dark energy. Thus one naturally arrives at the assumption of holographic dark energy with parameterizing its energy density as follows
 \begin{equation}
    \rho_{de}=3d^2 M_p^2 L^{-2}~, \label{i1}
 \end{equation}
with $d$ a positive constant parameter. Then the fractional energy density is given by
 \begin{equation}
    \Omega_{de}=\frac{\rho_{de}}{3M_p^2H^2}=\frac{d^2}{H^2L^2} ~. \label{i2}
 \end{equation}
with $H$ the Hubble parameter. Here the IR cut-off $L$ is regarded as the characteristic length scale of the universe.

There are a number of holographic dark energy models \cite{Li:2004rb, Cai:2007us, Wei:2007ty, Wei:2007xu,Gao:2007ep, Gao:2011,Huang:2012nz,Huang:2012gd} proposed by
choosing different characteristic length scales of the universe. It is noted that the same formula Eq.(\ref{vac}) may also be resulted from a different motivation
with the age of the universe \cite{Cai:2007us, Wei:2007ty} based on the metric fluctuation of Minkowski spacetime \cite{MF} and the corresponding quantum energy
density \cite{ED}, which follows from the quantum line fluctuation of spacetime \cite{LF}. The so-called new agegraphic dark energy model was yielded by taking the
conformal time of the universe without including the inflationary stage of the universe \cite{Wei:2007ty}, i.e. $L=\int^t \frac{dt'}{a(t')}$ with $a(t)$ the scale
factor of the universe, to be the IR cutoff of the universe. Due to the analytical feature that $\Omega_{de} \propto  a^2$ in matter-dominated epoch, the new
agegraphic dark energy model is a single-parameter model like the $\Lambda$CDM model. The best-fit analysis in \cite{Wei:2007xu} indicated that the model parameter
$n$ ( that corresponds to the parameter $d$ in our present notation, i.e., $n\equiv d$ ) is at the order of one, $n\equiv d\simeq O(1)$. A recent analysis was made
by considering the radiation contribution in the initial condition \cite{LZZ} and still yielded a similar result for the single parameter $n\equiv d\simeq O(1)$.
Such a model was extended in \cite{Sun:2011vg} to include the stage of inflation \cite{Guth:1980zm, Linde:1981mu, Albrecht:1982wi} so as to avoid the possible
divergence in the energy density caused by across zero transition of the conformal time during the universe expansion. Where the conformal time was redefined to be
$L=\left(\int^t \frac{dt'}{a(t')}+\delta\right)$ with $\delta$ a positive constant, obviously such a redefined conformal time becomes positive during the inflation.
Taking the single model parameter to be $n\equiv d\simeq O(1)$ yielded in \cite{Wei:2007xu} from the best-fit analysis without considering the inflation of
universe, then it was concluded in \cite{Sun:2011vg} that the resulting fraction of dark energy at the present epoch is too small $\Omega_{de0} \ll 1$  and
contradictory to the cosmological observations. The subscript "0" always indicates the present values of the corresponding quantities. Thus the new agegraphic dark
energy model proposed in \cite{Wei:2007ty,Wei:2007xu} was shown to be inconsistent with the inflationary universe. However, from Eqs.(\ref{i1}) and (\ref{i2}), it
is conceivable that the redefinition of conformal time with including the inflationary effect of the universe may change the behavior of the energy density and also
the fractional energy density of the holographic dark energy in the early universe. Therefore, the resulting best-fit value $n\equiv d\simeq O(1)$ in
\cite{Wei:2007xu} based on the analytical feature $\Omega_{de}\propto a^2$ during the matter-dominated epoch as well as its inconsistency with the inflationary
universe concluded in \cite{Sun:2011vg} should be reconsidered and studied carefully.

In the cosmological study, the conformal time is traditionally defined as the comoving horizon of the universe since the epoch $t_e$ at the end of the inflation,
i.e., $\tilde{\eta} = \int_{t_e}^t \frac{dt'}{a(t')}$, which may be yielded by subtracting off a very large primordial part $\eta_{\rm prim}$ generated by the
inflation from the total comoving horizon of the universe $\eta=\int_0^t \frac{dt'}{a(t')}$, namely, $\tilde{\eta} = \eta - \eta_{\rm prim}$. This is why the
conformal time during the inflation is thought to be negative and the usual conformal time of the universe has a beginning at $\tilde{\eta} = 0$. In general, it is
more reasonable to take the total comoving horizon to be correlated to the IR cutoff of the universe instead of the usual conformal time of universe, so that there
would be no longer divergence in the energy density when applying for the holographic principle.

Recently, we proposed in refs.\cite{Huang:2012nz,Huang:2012gd} an interesting holographic dark energy characterized by the conformal-age-like length (CHDE) with
including the epoch of inflation. Where the characteristic length $L$ of the universe is defined as $L=\frac{1}{a^4(t)}\int_0^t dt'~a^3(t') = \frac{1}{a^4(t)}\int
d\eta'~a^4(\eta')$, which is motivated from the four dimensional spacetime volume at cosmic time t of the flat Friedmann-Robertson-Walker (FRW) universe. As a
consequence, the possible divergence in the energy density can appropriately be avoided and meanwhile the very large primordial part of the characteristic length
$L$ of the universe generated by the inflation is associated with a power-law suppressing factor $(a_b/a_e)^4$, here $a_b$ and $a_e$ correspond to the scale factors
of the universe at the beginning and the end of the inflation, respectively. Thus the characteristic length $L$ of the universe has been shown to be dominated by
the conformal-age-like length $L\propto \frac{1}{Ha}$. To be more explicit, considering the epoch of universe dominated by the ambient constituent with constant
$w_m\geq-1$, we have approximately $H^2 \propto a^{-3(1+w_m)}$ from Fridemann equation, and obtain  $L=\left( \frac{a_b}{a} \right)^4
L_b+\frac{2}{3(3+w_m)}\left(\frac{1}{Ha}- \frac{1}{H_ba_b}\left( \frac{a_b}{a} \right)^4  \right)$, with $L_b=\frac{1}{a_b^4}\int_0^{t_b} dt'~a^3(t')$, here
subscript $b$ denotes the value of the corresponding quantity at the beginning of ambient-constituent-dominated epoch. Due to the expansion of the universe and
 $H^2 \propto a^{-3(1+w_m)}$ approximately, all the terms with factor $\left(a_b/a\right)^4$ can be ignored safely comparing to the term without such a factor. Thus
the fractional dark energy density is found to be $\Omega_{de}\simeq\frac94(3+w_m)^2d^2 a^2$ in the early time of universe when $a\ll 1$. It has been shown that the
CHDE model can be regarded as a single-parameter model like the $\Lambda$CDM model, so that the present fractional energy density $\Omega_{de}(a=1)$ can solely be
determined by solving the differential equation of $\Omega_{de}$ \cite{Huang:2012nz}. A systematic analysis and cosmological constraint from the recent cosmological
observations leads to $\Omega_{de}(a=1)=0.714$  and $d \simeq 0.232$ \cite{Huang:2012gd}.

In this paper, we are going to consider an alternative holographic dark energy characterized by the total comoving horizon $\eta$ of the universe (that may be
denoted in an abbreviation $\eta$HDE for convenience of mention below). We will show in detail how such an $\eta$HDE model is consistent with the inflationary
universe and  behaves like a cosmological constant. The paper is organized as follows: In Sec. II, we describe in detail the holographic dark energy characterized
by the total comoving horizon including the epoch of inflation; In Sec. III, we make a systematic cosmological analysis on $\eta$HDE and obtain consistent
cosmological constraints on $\eta$HDE. As a consequence, the $\eta$HDE is compatible with the cosmological constant, so that both the fine-tuning problem and the
coincidence problem can reasonably be understood with the inflationary universe and holographical principle; In Sec. IV, we come to our conclusions and remarks on
the $\eta$HDE model.

\section{Inflationary Universe and Holographic Dark Energy Model}

The total comoving horizon of the universe is defined as
 \begin{equation}
    \eta=\int_0^t \frac{c\ dt'}{a(t')}=\int_0^a \frac{d a'}{a'} \frac{1}{H'a'}  ~, \label{eta}
 \end{equation}
where $c=1$ is the speed of light in natural units, $a(t)$ is the scale factor, $H\equiv \dot{a}/a$ is the Hubble parameter and $``\cdot"$ is the derivative respect
to cosmic time $t$.

Note that the regions separated by the comoving distances greater than $\eta$ are not causally connected. Thus, it is reasonable to take the IR cutoff of the
universe $L$ to be less than the total comoving horizon,
\begin{equation}
L = \frac{1}{c_h} ~\eta~, \qquad c_h > 1
\end{equation}
where the constant $c_h$ can be regarded as a scale factor between the total comoving horizon and the characteristic length of universe.
With the above consideration, the energy density of corresponding holographic dark energy Eq.(\ref{vac}) can be rewritten as the following form
in terms of the total comoving horizon $\eta$
 \begin{equation}
    \rho_{de}=3d^2 M_p^2 L^{-2}= 3 (c_h d)^2 M_p^2 \eta^{-2}~, \label{rho}
 \end{equation}
with the combined parameter $c_h d$ as an independent model parameter. For convenience, we shall reexpress $c_h d$ to be a rescaled parameter denoted by the same
notation $d$,  i.e.,
\begin{equation}
c_hd \to d ~.
 \end{equation}
Then the fractional energy density for the holographic dark energy is defined as
 \begin{equation}
    \Omega_{de}=\frac{\rho_{de}}{3M_p^2H^2}=\frac{d^2}{H^2\eta^2} ~. \label{frho}
 \end{equation}
Here the rescaled parameter $d$ reflects the magnitude of the scale factor $c_h$ between the total comoving horizon and the characteristic length of universe $c_h =
\eta/L$. Obviously, the rescaled parameter $d$ can in general be much larger than one, $d\gg 1$ when $c_h \gg 1$.

The primordial part of the comoving horizon $\eta_{\rm prim}$ generated by inflation is given by
 \begin{equation}
    \eta_{\rm prim}=\int_{a_b}^{a_e} \frac{d a'}{a'} \frac{1}{H'a'}  ~,
 \end{equation}
where subscripts $b$ and $e$ denote the beginning and the end of inflation respectively. As the Hubble parameter $H$ is approximately constant in the slow-rolling
inflation model,  the primordial part of the comoving horizon $\eta_{\rm prim}$ is found to be
 \begin{equation}
 \eta_{\rm prim}\simeq \frac1{H_b a_b}-\frac1{H_e a_e} = \frac1{H_e a_e}\left( \frac{H_ea_e}{H_ba_b} -1 \right) \simeq \frac1{H_e a_e}\left( e^{H(t_e-t_b)} -1
 \right) ~.
\end{equation}
In order to solve the horizon problem \cite{Guth:1980zm}, the inflation is required to last enough time to make
 \begin{equation}
    \eta_{\rm prim}\simeq\frac1{H_b a_b}-\frac1{H_e a_e} > \frac1{H_0 a_0}  ~.\label{neq}
 \end{equation}

For the instantaneous reheating, the universe turns into the radiation-dominated epoch just after the end of the inflation, and then into the matter-dominated epoch
after the matter-radiation equality. In radiation-dominated and matter-dominated epoch, we have $H^2 \propto \rho_m \propto a^{-3(1+w_m)}$, with $w_m=\frac13$ for
radiation-dominated epoch and $w_m=0$ for matter-dominated epoch respectively. Thus, the part of comoving horizon generated in radiation- and matter-dominated epoch
after the inflation is given by
 \begin{equation}
    \tilde{\eta} =\int_{a_{e}}^{a} \frac{d a'}{a'} \frac{1}{H'a'}\simeq\frac{2}{1+3w_m}\left(\frac1{Ha}-\frac1{H_ea_e}\right)
    %\stackrel{a\gg a_{e}}{\simeq}
    \simeq \frac{2}{1+3w_m}\frac1{Ha} ~,
 \end{equation}
where the initial term $\frac{1}{H_ea_e}$ can be neglected for $a\gg a_e$ in a good approximation due to the fact that $\frac1{Ha}$ grows as $a^{(1+3w_m)/2}$ at
radiation- and matter-dominated epoch.

When ignoring the initial pre-inflation part, we then obtain the total comoving horizon of the universe at the radiation- and matter-dominated epoch
 \begin{equation}
    \eta = \eta_{\rm prim} + \tilde{\eta} =\eta_{\rm prim} + \int_{a_{e}}^{a} \frac{d a'}{a'} \frac{1}{H'a'} \simeq \eta_{\rm prim} + \frac{2}{1+3w_m}\frac1{Ha}  ~. \label{etam}
 \end{equation}
Taking the above total-comoving-horizon $\eta$ to characterize the holographic dark energy, we arrive at an $\eta$HDE model with the energy density and the
fractional energy density of dark energy given as follows
\begin{equation}
    \rho_{de}\simeq 3d^2 M_p^2 (\eta_{\rm prim}+ \tilde{\eta})^{-2}~,\quad  \Omega_{de}\simeq \frac{d^2}{H^2(\eta_{\rm prim} + \tilde{\eta})^2} \label{fracde}
 \end{equation}

Due to the fact that the universe has been dominated by radiation and matter until recently and $\frac1{Ha}$ scales as $a^{(1+3w_m)/2}$ in radiation- and
matter-dominated epoch, we have $\frac1{H_0 a_0} \gg \frac1{H a}$ when $a_e \ll a \ll a_0$. Combining with Eq.(\ref{neq}) and (\ref{etam}), we yield that at the
radiation- and matter-dominated epoch the comoving horizon is approximately a constant with $\eta \simeq \eta_{\rm prim}$, and the holographic dark energy behaves
like a cosmological constant with approximately constant energy density
 \begin{equation}
    \rho_{de}\simeq 3d^2 M_p^2 \eta_{\rm prim}^{-2}~. \label{rhoi}
 \end{equation}
Correspondingly, the fractional energy density of dark energy in the early universe when $H^2 \propto \rho_m \propto a^{-3(1+w_m)}$, scales as
 \begin{equation}
    \Omega_{de}\simeq \frac{d^2}{H^2\eta_{\rm prim}^2} \propto\frac{d^2}{\eta_{\rm prim}^2} a^{3(1+w_m)} ~.  \label{frhoi}
 \end{equation}
It is seen that in radiation- and matter-dominated epoch the fraction of dark energy scales as $\Omega_{de} \propto a^{3(1+w_m)}$ in the $\eta$HDE model, which
distinguishes from the new agegraphic dark energy model in which the fraction of dark energy scales as $\Omega_{de} \propto a^2$.

To further investigate the $\eta$HDE in recent cosmic acceleration, it needs a concrete cosmological model. Considering the holographic dark energy and matter with
equation of state $w_m=0$ ( $w_m$ will be kept in derivations ) in a flat Friedmann-Robertson-Walker~(FRW) universe, we have Friedmann equation
 \begin{equation}
    3M_p^2H^2=\rho_m +\rho_{de}  ~,
 \end{equation}
or
 \begin{equation}
    \Omega_{de}+\Omega_{m}=1  ~,   \label{fri}
 \end{equation}
in fractional energy densities. Where $ \Omega_{m}=\frac{\rho_{m}}{3M_p^2H^2}$ is the fractional energy density of matter. When each energy component is
conservative respectively, we have conservation equations
 \begin{equation}
    \dot{\rho}_{i}+3H(1+w_{i})\rho_{i}=0  ~ \label{ceq}
 \end{equation}
for $i=m,\, de$. From Eqs.(\ref{eta}), (\ref{rho}), (\ref{frho}) and (\ref{ceq}), we obtain the EoS of the dark energy
 \begin{equation}
    w_{de}=-1+\frac2{3d}\frac{\sqrt{\Omega_{de}}}a~.\label{wde}
 \end{equation}
Substituting the definition of $\Omega_{de}$ Eq.(\ref{frho}) into $w_{de}$ and using Eq.(\ref{etam}),  we have
 \begin{equation}
    w_{de}=-1+\frac2{3}\frac{\frac1{Ha}}{\eta}\simeq -1 + \frac2{3}\frac{\frac1{Ha}}{\eta_{\rm prim} +\frac{2}{1+3w_m} \frac1{Ha} } ~.
 \end{equation}
which indicates that $w_{de}\simeq -1$ in the early universe when $\eta_{\rm prim} \gg \frac1{Ha}$, namely, the dark energy behaves like a cosmological constant in
this case.

The conservation of matter results in $\rho_m=C_1 a^{-3(1+w_m)}$, with $C_1$ a constant proportionality coefficient. Combining with the definition of fractional
energy densities and the Friedmann equation, it is not difficult to get
 \begin{equation}
 \frac1{Ha}= \frac1{ \scriptstyle \sqrt{\frac{C_1}{3M_p^2}}} \sqrt{a^{(1+3w_m)}(1-\Omega_{de})}~.\label{ha}
 \end{equation}
While from Eqs.(\ref{eta}) and (\ref{frho}), we have
 \begin{equation}
    \int_0^a \frac{d a'}{a'} \frac{1}{H'a'}=\frac{d }{\sqrt{\Omega_{de}}H } ~.
 \end{equation}
Substituting Eq.(\ref{ha}) into above equation and taking derivative with respect to $a$ in both sides, we get the differential equation of motion for $\Omega_{de}$
 \begin{equation}
 \frac{d\Omega_{de}}{da}=\frac{\Omega_{de}}{a}(1-\Omega_{de})\left(3w_m+3-\frac2d \frac{\sqrt{\Omega_{de}}}a \right)~.\label{ode}
 \end{equation}

It is noted that the energy density Eq.(\ref{rho}), the fractional energy density Eq.(\ref{frho}), the EoS Eq.(\ref{wde}) and the differential equation
Eq.(\ref{ode}) are all invariant under the transformation $a \to \frac{a}{a_0}, ~ d \to d a_0$. Namely, after performing such a transformation, and taking $a_0$ to
be the present scale factor of the universe by setting $a_0=1$, all the expressions keep the same forms. From now on, we adopt that the parameter $d$ has absorbed a
factor $a_0$ and set $a_0=1$.

From the derivations it is interesting to show that the EoS Eq.(\ref{wde}) and the differential equation Eq.(\ref{ode}) are invariant under the translation $\eta
\to \eta+ \eta_{\delta}$, where $\eta_{\delta}$ is a constant length scale. However, the energy density and fractional energy density will change under such a
translation. Therefore, for self-consistency, any approximate solution derived from the differential equation Eq.(\ref{ode}) shall be checked with the fractional energy density Eq.(\ref{frho}) by a direct calculation.

Referring to Eq.(\ref{frhoi}) and setting $w_m=0$ there, it is seen that $\Omega_{de}\propto a^{3}$ in the matter-dominated epoch when $a \ll 1$, which can consistently be obtained from the equation of motion for $\Omega_{de}$ in Eq.(\ref{ode}) under the limit $1-\Omega_{de} \simeq 1$ when $a \ll 1$. This can easily be checked by noticing the definition of $\Omega_{de}$ in Eq.(\ref{fracde}), in the early universe when $\eta_{\rm prim} \gg \frac1{Ha}$, the last factor in the differential equation Eq.(\ref{ode}) is found to be
 \begin{equation}
    \frac1d\frac{\sqrt{\Omega_{de}}}a=\frac{\frac1{Ha}}\eta=\frac{\frac1{Ha}}{\eta_{\rm prim} +\frac{2}{1+3w_m} \frac1{Ha}} \ll 1
 \end{equation}
which is negligible small in comparison with the constant factor $3(1+w_m)$ in Eq.(\ref{ode}). In this case, the differential equation Eq.(\ref{ode}) is approximately given by
 \begin{equation}
 \frac{d\Omega_{de}}{da}\simeq \frac{\Omega_{de}}{a}3(w_m+1)~.\label{ode1}
 \end{equation}
which leads to the consistent approximate solution $\Omega_{de}\propto a^{3}$ in the matter-dominated epoch with $w_m=0$ and $a\ll 1$.

Note that the analytical result in new agegraphic dark energy model $\Omega_{de} \propto a^2$ is
inconsistent with the direct calculation, although it is also the approximate solution of Eq.(\ref{ode}) under the limit $1-\Omega_{de} \simeq 1$ when $a \ll 1$.
However, if we take the conformal time defined as $\eta-\eta_{\rm prim}$ by subtracting off the primordial part $\eta_{\rm prim}$ from the total comoving horizon to
be the IR cutoff, then we can show that $\Omega_{de} \simeq \frac{d^2}4 (3w_m+1)^2 a^2$ is the consistent approximate solution by taking the conformal time
$\eta-\eta_{\rm prim}\simeq\frac{2}{1+3w_m}\frac1{Ha}$ in the radiation- and matter-dominated epoch from Eq.(\ref{etam}). Therefore, the redefinition will change the
behavior of $\Omega_{de}$ in early universe. The analytical property of $\Omega_{de}$ in the limit $a\ll1$ has practically been realized in the CHDE model with
including the inflationary epoch of universe \cite{Huang:2012nz,Huang:2012gd}, where $\Omega_{de} \simeq \frac{9d^2}4 (w_m+3)^2 a^2$ in the early universe.

\section{Cosmological Analysis on $\eta$HDE and Insights to Cosmological Constant and Coincidence Problem}

In the $\eta$HDE model, there are in general two model parameters, since the proportionality coefficient of the analytical approximate solution $\Omega_{de}\propto
a^{3}$ in matter-dominated epoch can not be determined only by the parameter $d$, which distinguishes from the new agegraphic dark energy model. In the practical
calculation, we would like to choose two parameters as the usual parameter $d$ and the present fractional energy density of matter $\Omega_{m0}$. We are going to
use the Union2 compilation of 557 supernova Ia (SNIa) data \cite{Amanullah:2010vv}, the parameter $A$ \cite{Eisenstein:2005su} from BAO measurements and the shift
parameter $R$ \cite{Bond:1997wr} from CMB measurements \cite{Komatsu:2010fb} to perform the best-fit analysis. The analysis method for the observational data is
given in Appendix A. The best-fit result gets $\chi^2_{\rm min}=542.915$ for the best-fit parameters $d=209.86$ and $\Omega_{m0}=0.270$. As a comparison, we also
fit the $\Lambda$CDM model to the same observational data, and find that the minimal $\chi^2_{\Lambda {\rm CDM}}=542.919$ for the best fit parameter
$\Omega_{m0}=0.270$.

\begin{figure}
\centerline{\includegraphics[width=13cm, height=6cm]{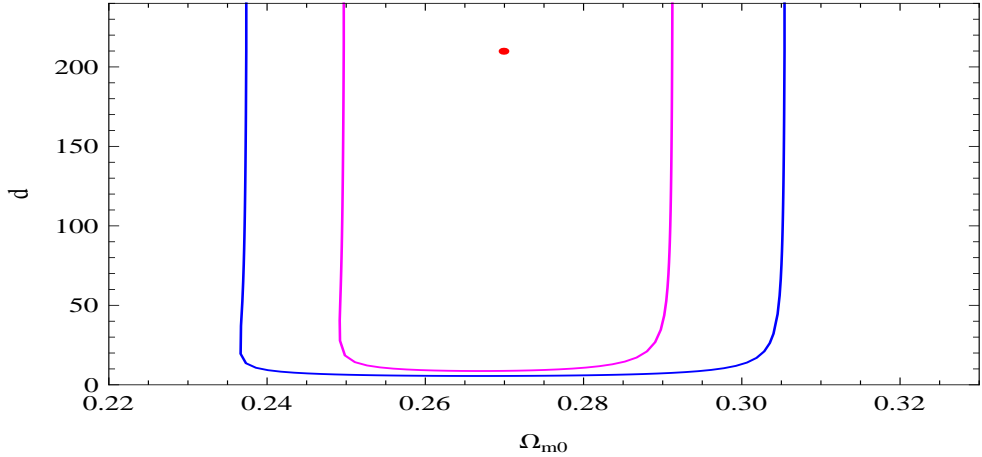}} \caption{\label{fit}  Probability contours at 68.3\% and 95.4\% confidence levels  of the holographic
dark energy model characterized by the total comoving horizon; the best-fit values are $d=209.86$ and $\Omega_{m0}=0.270$.}
\end{figure}

In Fig.[\ref{fit}], we plot probability contours at 68.3\% and 95.4\% confidence levels of the two parameters. It can be seen that the constraint on parameter $d$
is very weak, i.e. $d \geq 8.68$ at 1$\sigma$ level. This means that as long as the rescaled parameter $d$ is taken to be appropriately large $d \agt O(10)$, namely
the characteristic length scale of universe is much smaller than the total comoving horizon of universe, there is only one effective parameter $\Omega_{m0}$ in the
model just like the $\Lambda$CDM model. The reason why the parameter $d$ is not bounded above can be enlightened from the differential equation of motion for
$\Omega_{de}$. In the limit $d \to \infty$, Eq.(\ref{ode}) reduces to
 \begin{equation}
    \frac{d\Omega_{de}}{da}=\frac{\Omega_{de}}{a}(1-\Omega_{de})(3w_m+3)~,
 \end{equation}
which is just the the differential equation of motion for $\Omega_{\Lambda}$ in the $\Lambda$CDM model. Since $\chi^2_{\Lambda {\rm CDM}}=542.919$ is  almost the
same to that of the $\eta$HDE model, the constraint on parameter $d$ can not be bounded above. This shows that the $\eta$HDE model behaves just like the
$\Lambda$CDM model once the value of  $d$ is taken to be large enough.

\begin{figure}
\centerline{
\includegraphics[width=8cm,height=8cm]{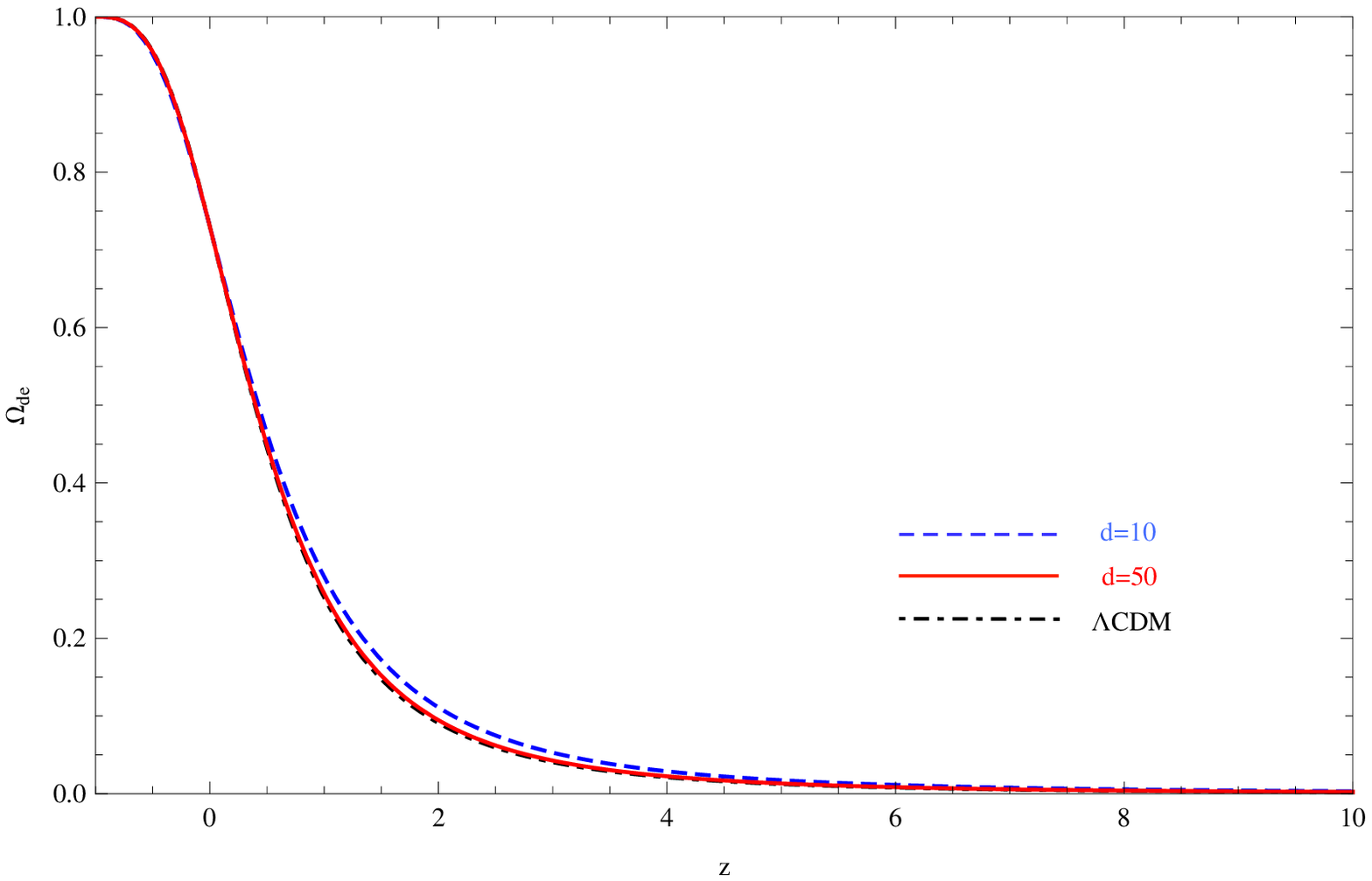}
\includegraphics[width=8cm,height=8cm]{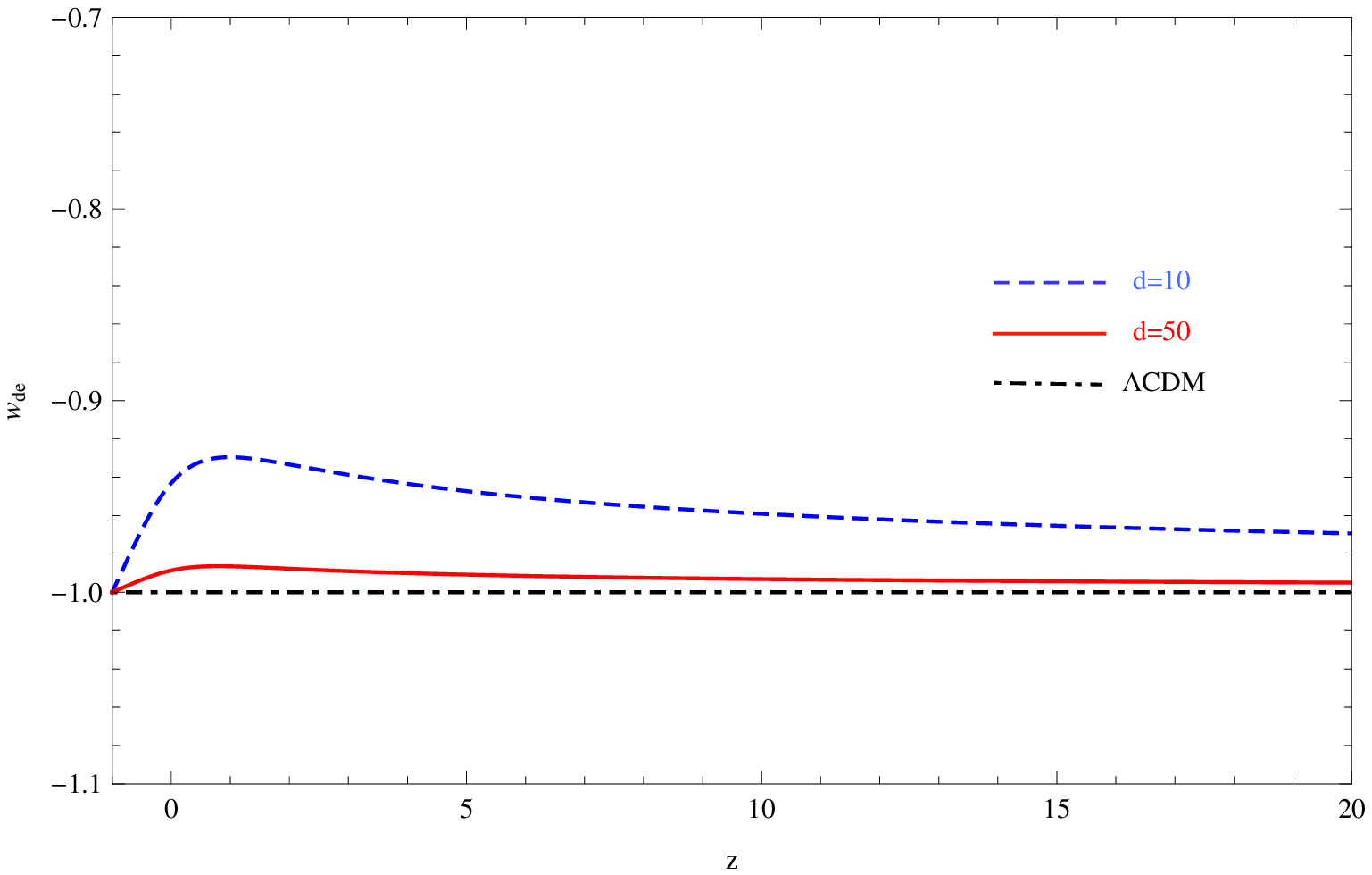}
}
 \centerline{
\includegraphics[width=14cm,height=8cm]{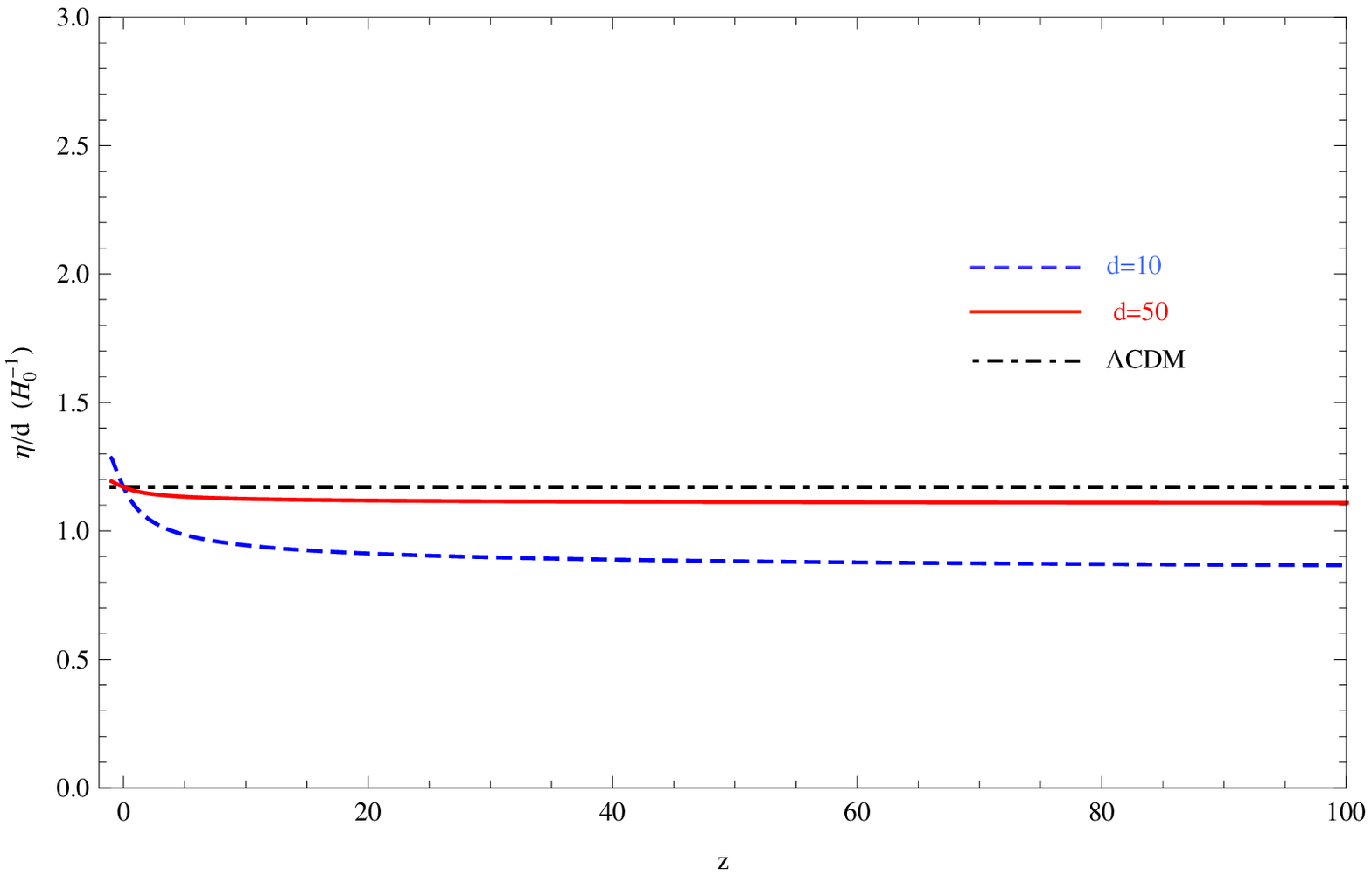}
} \caption{\label{fig2} The evolutions of the fractional energy density $\Omega_{de}$ , the EoS of dark energy $w_{de}$ and the ratio $\eta/d$ ( in units of current
horizon $H_0^{-1}$) corresponding to different $d$, where $\Omega_{m0}=0.270$ is fixed.}
\end{figure}

In Fig.[\ref{fig2}], we show the evolutions of the fractional energy density, the EoS of dark energy and the ratio $\eta/d$ ( in units of current horizon $H_0^{-1}$
) corresponding to different $d$ by fixing $\Omega_{m0}=0.270$. It is seen that even the parameter $d$ takes value at the order $d\simeq O(10)$, the $\eta$HDE
behaves like a cosmological constant. The corresponding cosmological constant is given by the ratio $\eta/d$ which is at the order of the current horizon size of
universe. By comparing the ratio $\eta/d$ at high redshift to the approximate value $\eta_{\rm prim}/d$, we have $\eta_{\rm prim}\sim O(dH_0^{-1})$.

On the other hand, when the parameter $\eta_{\rm prim}$ is known from inflation and the parameter $d$ is given, the dark energy density $\rho_{de}$ in the early
universe is obtained approximately from Eq.(\ref{rhoi}). Then, we may use this condition, the friedmann equation and the differential equation of $\Omega_{de}$ to
predict the present $d/\eta$, which gives the corresponding cosmological constant. Based on this analysis, the small value of the corresponding cosmological
constant is due to the very large $\eta_{\rm prim}$ by lasting many e-folds in the inflationary epoch when the parameter $d$ is taken to be a normal value. It is
interesting to notice that as both the values of the corresponding cosmological constant and the radiation energy density at the end of inflation are determined by
the inflationary universe in the present scenario, the coincidence problem can appropriately be solved by the inflationary universe.

\section{Conclusions and Remarks}

We have proposed a consistent holographic dark energy characterized by the total comoving horizon of universe ($\eta$HDE) to explain the acceleration of present
universe. It has been shown that due to the very large primordial part of comoving horizon $\eta_{\rm prim}$ generated by the inflation, the $\eta$HDE behaves like
a cosmological constant.

It has been seen that the $\eta$HDE model described in this paper distinguishes from the new agegraphic dark energy model \cite{Wei:2007ty} in which the subtracted
comoving horizon $\tilde{\eta} =\eta-\eta_{\rm prim}$ was taken to be the conformal age as defined widely in literature, this may be seen from the approximate
solution $\Omega_{de} \propto a^2$ used in \cite{Wei:2007ty} for the matter-dominated epoch. A redefined conformal age $\tilde{\eta} + \delta$ with the positive
constant $\delta$ was introduced in \cite{Sun:2011vg} to remove the possible divergence at $\tilde{\eta} = 0$ by extending the new agegraphic dark energy model to
include the inflation stage of the universe, while the study in ref.\cite{Sun:2011vg} arrived at the conclusion that $\Omega_{de0} \ll 1$ and the new agegraphic
dark energy cannot become dominated at present time. The main reason is that the condition $n\equiv d \sim O(1)$ on which the argument based in \cite{Sun:2011vg}
was derived from the best-fit analysis on the new agegraphic dark energy model \cite{Wei:2007ty},  where the approximate solution $\Omega_{de} \propto a^2$ in
matter-dominated epoch was used as the initial condition to solve the differential equation of motion for $\Omega_{de}$. Just as mentioned in the previous section
that the large primordial part $\eta_{\rm prim}$ consistently makes $\Omega_{de} \propto a^{3}$ in matter-dominated epoch rather than $\Omega_{de} \propto a^2$. As
a consequence, the constraint on the parameter $d$ from the cosmological observations becomes very weak, i.e. $d \geq 8.68$ at 1$\sigma$ level. Namely, as long as
the total comoving horizon of universe $\eta$ is much larger than the characteristic length scale of the universe, i.e., $c_h=\eta/L \gg 1$, there is only one
effective parameter in the $\eta$HDE model which behaves like the $\Lambda$CDM model. Therefore, it is appropriate to have a large rescaled parameter $d\sim O(10)$
for arriving at the fractional dark energy density $\Omega_{de0} \simeq 0.73$ to be consistent with the current cosmological observations.

As shown in \cite{Li:2009bn}, the new agegraphic dark energy model \cite{Wei:2007ty} fitting to cosmological observations is worse than the $\Lambda$CDM model. In
order to be cooperative with the inflation \cite{Guth:1980zm, Linde:1981mu, Albrecht:1982wi} which is an important idea in modern cosmology, it is more reasonable to take the total comoving horizon or total conformal age including the epoch of inflation to be associated with the characteristic length scale. Thus the new agegraphic dark energy model with the
analytic feature $\Omega_{de} \simeq \frac{d^2}4 (3w_m+1)^2 a^2$ at $a\ll 1$ considered in \cite{Wei:2007ty} appears to be inappropriate for a complete description.
Actually, the new agegraphic dark energy model was based on the metric fluctuation of Minkowski spacetime \cite{MF} and the corresponding quantum energy density
\cite{ED}, it is also natural to take the total conformal age $\eta/c$ as the characteristic age scale in the model. Although these two models have the same formula, while they are resulted from different physical considerations. From the effective quantum field theory and the holographic principle,  it may lead us to have a deeper understanding on the physical origin of dark energy.

It is  noticed that the EoS in the $\eta$HDE model always has $w_{de} \agt -1$ which slightly deviates from the EoS $w_{de} = -1$ for the $\Lambda$CDM model, but it
is different from the CHDE model proposed recently in \cite{Huang:2012nz,Huang:2012gd}, where the EoS has been shown to cross from $w_{de}>-1$ to $w_{de}<-1$ during
the universe expansion and lead the present EoS of the CHDE to be smaller than $-1$. Also the CHDE model gives a slightly larger fraction of matter
\cite{Huang:2012gd}.

In conclusion, the $\eta$HDE model provides a natural and consistent explanation on the observed accelerated expansion of the present universe. The $\eta$HDE model
is compatible with the $\Lambda$CDM model as the $\eta$HDE behaves like a cosmological constant. But unlike the cosmological constant, the $\eta$HDE has no longer a
serious fine-tuning problem and the coincidence problem as both of them can be understood reasonably with the inflationary universe and holographic principle.

\section*{Acknowledgements}
We would like to thank R.~G. Cai, Q.~G.~Huang, Miao Li and J.~X. Lu for useful discussions. The author (Z.P.H) would like to thank M. Q. Huang and M. Zhong for
their helpful support. This work is supported in part by the National Basic Research Program of China (973 Program) under Grants No. 2010CB833000; the National
Nature Science Foundation of China (NSFC) under Grants No. 10975170, 10975184, 10947016.

\appendix

\section{ Analysis method for the observational data}

In this appendix, we present the method used for the best-fit analysis on the observational data from Type Ia Supernovae (SNIa), Baryon Acoustic Oscillations (BAO),
Cosmic Microwave Background (CMB) radiation.

For the three independent observations, the likelihood function of a joint analysis is in general given by
 \begin{eqnarray}
  \cal{L }&=& \cal{L }_{\rm SN} \times \cal{L }_{\rm BAO} \;\; \times \cal{L }_{\rm CMB}  \nonumber \\
    &=& \exp[-(\tilde{\chi}^2_{\rm SN}+\chi_{\rm BAO}^2+\chi_{\rm CMB}^2)/2]~.
\end{eqnarray}
A maximal $\cal{L }$ for the model parameters corresponds to a minimal $\chi^{2}=\tilde{\chi}_{\rm SN}^{2}+\chi_{\rm BAO}^{2}+\chi_{
\rm CMB}^{2}$. In the following, we present the calculation for the various  $\chi^{2}_i$ of each observational
data set.  A description for the analysis method may also be seen in \cite{Geng:2011ka,Bamba:2010wb,Yang:2010xq}.

\subsection{Type Ia Supernovae (SN Ia)}

The SNIa observations give the information on the luminosity distance $D_{L}$ as a function of the redshift $z$. The distance modulus is theoretically defined as
\begin{equation}
\mu_{\rm th}(z_i)\equiv 5 \log_{10} {D_L(z_i)} +\mu_0   ~,
\end{equation}
with $\mu_{0}\equiv42.38-5\log_{10}h$, and $h \equiv H_{0}/100/[{\rm km} \, {\rm sec}^{-1} \, {\rm Mpc}^{-1}]$. The Hubble-free luminosity distance has the following form for the flat universe
\begin{equation}
 D_{L}(z)=\left(1+z\right)\int_{0}^{z}\frac{dz'}{E(z')} ~,
\end{equation}
with $E(z) \equiv H(z)/H_{0}$.

The $\chi^2$ for the SNIa data is given by
\begin{equation}
\chi^2_{\rm SN}=\sum\limits_{i}{[\mu_{\rm obs}(z_i)-\mu_{\rm th}(z_i)]^2\over \sigma_i^2} ~, \label{ochisn}
\end{equation}
where $\mu_{\rm obs}(z_i)$ and $\sigma_i$ are the observed quantity and the corresponding 1$\sigma$ error of distance modulus for each supernova, respectively.
Applying the approach given in \cite{Perivolaropoulos:2004yr}, the $\chi^2_{\rm SN}$ can be expanded with respect to $\mu_0$ as follows
\begin{equation}
\chi^2_{\rm SN}=A-2\mu_0 B+\mu_0^2 C ~, \label{ochisn2}
\end{equation}
with
\begin{equation}
A=\sum\limits_{i}{[\mu_{\rm obs}(z_i)-\mu_{\rm th}(z_i;\mu_{0}=0)]^2\over \sigma_i^2}  ~,
\end{equation}
\begin{equation}
B=\sum\limits_{i}{\mu_{\rm obs}(z_i)-\mu_{\rm th}(z_i;\mu_{0}=0)\over \sigma_i^2} ~,
\end{equation}
\begin{equation}
C=\sum\limits_{i}{1\over \sigma_i^2} ~.
\end{equation}
Thus the minimum of $\chi_{\mathrm{SN}}^{2}$ with respect to $\mu_{0}$ is found to be
\begin{equation}
\tilde{\chi}^2_{\rm SN}=A-\frac{B^2}C ~. \label{tchi2sn}
\end{equation}
which is used in our best-fit analysis to the $\chi^{2}$ minimization by using the Supernova Cosmology Project (SCP) Union2 compilation, which contains 557
supernovae~\cite{Amanullah:2010vv} with the range of the redshift $z=0.015-1.4$.

\subsection{Baryon Acoustic Oscillations (BAO)}

 From the distribution of SDSS luminous red galaxies, the quantity $A$ via the measurement of the BAO peak is defined as \cite{Eisenstein:2005su}
 \begin{equation}
 A\equiv \Omega_{m0}^{1/2} E(z_{b})^{-1/3}\left({1\over z_{b}}\int_0^{z_{b}}{dz'\over E(z')}\right)^{2/3}~,
 \end{equation}
at the redshift $z_{b}=0.35$. Where $A$ has been found to be $0.469\ (n_s/0.98)^{-0.35}\pm0.017$ \cite{Eisenstein:2005su}. $n_s$ is the scalar spectral index and
has been updated from the WMAP7 \cite{Komatsu:2010fb} to be $n_s=0.963$. The $\chi^2$ of the BAO data is given by:
 \begin{equation}
 \chi^2_{\rm BAO} = \frac{(A-A_{\rm obs})^2}{0.017^2} ~.  \label{chi2bao}
 \end{equation}

\subsection{The Shift Parameter from Cosmic Microwave Background (CMB)}

The shift parameter $R$, which relates the angular diameter distance to the last scattering surface and the comoving size of the sound horizon at $z_{*}$ as well as
the angular scale of the first acoustic peak in CMB power spectrum of temperature fluctuations, is given by \cite{Bond:1997wr}
 \begin{equation}
 R\equiv \Omega_{m0}^{1/2}\int_0^{z_{*}}{dz'\over E(z')}~,
 \end{equation}
where the redshift of the recombination has been updated in the WMAP7 \cite{Komatsu:2010fb} to be $z_{*}=1091.3$. The shift parameter $R$ has been updated to be
$1.725\pm0.018$~\cite{Komatsu:2010fb}. The $\chi^{2}$ of the CMB data is given by:
 \begin{equation}
 \chi^2_{\rm CMB} =\frac{(R-1.725)^2}{0.018^2} ~. \label{chi2cmb}
 \end{equation}

%----------------------------------------------------------------------------------


\vspace*{0.5cm}
\begin{thebibliography}{99}

%\cite{Riess:1998cb}
\bibitem{Riess:1998cb}
  A.~G.~Riess {\it et al.}  [Supernova Search Team Collaboration],
  %``Observational Evidence from Supernovae for an Accelerating Universe and a
  %Cosmological Constant,''
  Astron.\ J.\  {\bf 116}, 1009 (1998)
  [arXiv:astro-ph/9805201].
  %%CITATION = ANJOA,116,1009;%%

%\cite{Perlmutter:1998np}
\bibitem{Perlmutter:1998np}
  S.~Perlmutter {\it et al.}  [Supernova Cosmology Project Collaboration],
  %``Measurements of Omega and Lambda from 42 High-Redshift Supernovae,''
  Astrophys.\ J.\  {\bf 517}, 565 (1999)
  [arXiv:astro-ph/9812133].
  %%CITATION = ASJOA,517,565;%%

%\cite{Spergel:2003cb}
\bibitem{Spergel:2003cb}
  D.~N.~Spergel {\it et al.}  [WMAP Collaboration],
  %``First Year Wilkinson Microwave Anisotropy Probe (WMAP) Observations:
  %Determination of Cosmological Parameters,''
  Astrophys.\ J.\ Suppl.\  {\bf 148}, 175 (2003)
  [arXiv:astro-ph/0302209].
  %%CITATION = APJSA,148,175;%%

%\cite{Tegmark:2003ud}
\bibitem{Tegmark:2003ud}
  M.~Tegmark {\it et al.}  [SDSS Collaboration],
  %``Cosmological parameters from SDSS and WMAP,''
  Phys.\ Rev.\  D {\bf 69}, 103501 (2004)
  [arXiv:astro-ph/0310723].
  %%CITATION = PHRVA,D69,103501;%%


%\cite{Copeland:2006wr}
\bibitem{Copeland:2006wr}
  E.~J.~Copeland, M.~Sami and S.~Tsujikawa,
  %``Dynamics of dark energy,''
  Int.\ J.\ Mod.\ Phys.\  D {\bf 15}, 1753 (2006)
  [arXiv:hep-th/0603057].
  %%CITATION = IMPAE,D15,1753;%%


%\cite{Li:2011sd}
\bibitem{Li:2011sd}
  M.~Li, X.~-D.~Li, S.~Wang, Y.~Wang,
  %``Dark Energy,''
  Commun.\ Theor.\ Phys.\  {\bf 56}, 525-604 (2011)
  [arXiv:1103.5870 [astro-ph.CO]].



%\cite{Weinberg:1988cp}
\bibitem{Weinberg:1988cp}
  S.~Weinberg,
  %``The Cosmological Constant Problem,''
  Rev.\ Mod.\ Phys.\  {\bf 61}, 1-23 (1989);
%
%\cite{Weinberg:2000yb}
%\bibitem{Weinberg:2000yb}
%  S.~Weinberg,
  %``The Cosmological constant problems,''
[arXiv:astro-ph/0005265].



%\cite{'tHooft:1993gx}
\bibitem{'tHooft:1993gx}
  G.~'t Hooft,
  %``Dimensional reduction in quantum gravity,''
  arXiv:gr-qc/9310026.


%\cite{Susskind:1994vu}
\bibitem{Susskind:1994vu}
  L.~Susskind,
  %``The World as a hologram,''
  J.\ Math.\ Phys.\  {\bf 36}, 6377-6396 (1995)
  [arXiv:hep-th/9409089].


%\cite{Cohen:1998zx}
\bibitem{Cohen:1998zx}
  A.~G.~Cohen, D.~B.~Kaplan, A.~E.~Nelson,
  %``Effective field theory, black holes, and the cosmological constant,''
  Phys.\ Rev.\ Lett.\  {\bf 82}, 4971-4974 (1999)
  [arXiv:hep-th/9803132].


%\cite{Li:2004rb}
\bibitem{Li:2004rb}
  M.~Li,
  %``A Model of holographic dark energy,''
  Phys.\ Lett.\ B  {\bf 603}, 1 (2004)
  [arXiv:hep-th/0403127].

%\cite{Cai:2007us}
\bibitem{Cai:2007us}
  R.~-G.~Cai,
  %``A Dark Energy Model Characterized by the Age of the Universe,''
  Phys.\ Lett.\ B  {\bf 657}, 228-231 (2007)
  [arXiv:0707.4049 [hep-th]].

%\cite{Wei:2007ty}
\bibitem{Wei:2007ty}
  H.~Wei, R.~-G.~Cai,
  %``A New Model of Agegraphic Dark Energy,''
  Phys.\ Lett.\ B {\bf 660}, 113-117 (2008)
  [arXiv:0708.0884 [astro-ph]].

%\cite{Wei:2007xu}
\bibitem{Wei:2007xu}
  H.~Wei and R.~-G.~Cai,
  %``Cosmological Constraints on New Agegraphic Dark Energy,''
  Phys.\ Lett.\ B {\bf 663}, 1 (2008)
  [arXiv:0708.1894 [astro-ph]].
%%CITATION = ARXIV:0708.1894;%%



%\cite{Gao:2007ep}
\bibitem{Gao:2007ep}
  C.~Gao, X.~Chen, Y.~-G.~Shen,
  %``A Holographic Dark Energy Model from Ricci Scalar Curvature,''
  Phys.\ Rev.\  D {\bf 79}, 043511 (2009)
  [arXiv:0712.1394 [astro-ph]].

%\cite{Gao:2011}
\bibitem{Gao:2011}
  C.~Gao,
  %``Scalar Field, Four Dimensional Spacetime Volume and the Holographic Dark Energy,''
  arXiv:1108.5827 [gr-qc].


%\cite{Huang:2012nz}
\bibitem{Huang:2012nz}
  Z.~P.~Huang and Y.~L.~Wu,
  %``Holographic Dark Energy Model Characterized by the Conformal-age-like Length,''
  arXiv:1202.2590 [hep-th], to be published.
  %%CITATION = ARXIV:1202.2590;%%


  %\cite{Huang:2012gd}
\bibitem{Huang:2012gd}
  Z.~P.~Huang and Y.~L.~Wu,
  %``Cosmological Constraint and Analysis on Holographic Dark Energy Model
  %Characterized by the Conformal-age-like Length,''
  arXiv:1202.3517 [astro-ph.CO].
  %%CITATION = ARXIV:1202.3517;%%


\bibitem{MF}
M. Maziashvili, Int. J. Mod. Phys. D 16, 1531 (2007) [gr-qc/0612110]; \\
M. Maziashvili, Phys. Lett. B 652, 165 (2007) [arXiv:0705.0924].
\bibitem{ED} N. Sasakura, Prog. Theor. Phys. 102, 169 (1999) [hep-th/9903146]; \\
Y. J. Ng and H. Van Dam, Mod. Phys. Lett. A 9, 335 (1994); \\
Y. J. Ng and H. Van Dam, Mod. Phys. Lett. A 10, 2801 (1995);\\
S. Lloyd and Y. J. Ng, Sci. Am. 291, 52 (2004);\\
W. A. Christiansen, Y. J. Ng and H. van Dam, Phys. Rev. Lett. 96, 051301 (2006) [gr-qc/0508121];\\
M. Arzano, T. W. Kephart and Y. J. Ng, Phys. Lett. B 649, 243 (2007) [gr-qc/0605117].

\bibitem{LF}
F. K¡äarolyh¡äazy, Nuovo Cim. A 42, 390 (1966); \\
F. K¡äarolyh¡äazy, A. Frenkel and B. Luk¡äacs, in Physics as Natural Philosophy, edited by A. Simony and
H. Feschbach, MIT Press, Cambridge, MA (1982); \\
F. K¡äarolyh¡äazy, A. Frenkel and B. Luk¡äacs, in Quantum Concepts in Space and Time, edited by R. Penrose
and C. J. Isham, Clarendon Press, Oxford (1986).

\bibitem{LZZ}

Y.~H.~ Li, J.~F.~ Zhang and X.~ Zhang, arXiv: 1201.5446 [gr-qc].

%\cite{Sun:2011vg}
\bibitem{Sun:2011vg}
  C.~-Y.~Sun, R.~-H.~Yue,
  %``Inflation and New Agegraphic Dark Energy,''
  Phys.\ Rev.\ D  {\bf 83}, 107302 (2011)
  [arXiv:1101.1748 [hep-th]].


%\cite{Guth:1980zm}
\bibitem{Guth:1980zm}
  A.~H.~Guth,
  %``The Inflationary Universe: A Possible Solution to the Horizon and Flatness Problems,''
  Phys.\ Rev.\ D  {\bf 23}, 347-356 (1981).

%\cite{Linde:1981mu}
\bibitem{Linde:1981mu}
  A.~D.~Linde,
  %``A New Inflationary Universe Scenario: A Possible Solution of the Horizon, Flatness, Homogeneity, Isotropy and Primordial Monopole Problems,''
  Phys.\ Lett.\ B {\bf 108}, 389-393 (1982).

%\cite{Albrecht:1982wi}
\bibitem{Albrecht:1982wi}
  A.~Albrecht, P.~J.~Steinhardt,
  %``Cosmology for Grand Unified Theories with Radiatively Induced Symmetry Breaking,''
  Phys.\ Rev.\ Lett.\  {\bf 48}, 1220-1223 (1982).




%\cite{Amanullah:2010vv}
\bibitem{Amanullah:2010vv}
  R.~Amanullah {\it et al.},
  %``Spectra and Light Curves of Six Type Ia Supernovae at 0.511 < z < 1.12 and
  %the Union2 Compilation,''
  Astrophys.\ J.\  {\bf 716}, 712 (2010)
  [arXiv:1004.1711 [astro-ph.CO]].
  %%CITATION = ASJOA,716,712;%%



%\cite{Eisenstein:2005su}
\bibitem{Eisenstein:2005su}
  D.~J.~Eisenstein {\it et al.}  [SDSS Collaboration],
  %``Detection of the Baryon Acoustic Peak in the Large-Scale Correlation
  %Function of SDSS Luminous Red Galaxies,''
  Astrophys.\ J.\  {\bf 633}, 560 (2005)
  [arXiv:astro-ph/0501171].
  %%CITATION = ASJOA,633,560;%%

%\cite{Bond:1997wr}
\bibitem{Bond:1997wr}
  J.~R.~Bond, G.~Efstathiou and M.~Tegmark,
  %``Forecasting Cosmic Parameter Errors from Microwave Background Anisotropy
  %Experiments,''
  Mon.\ Not.\ Roy.\ Astron.\ Soc.\  {\bf 291}, L33 (1997)
  [arXiv:astro-ph/9702100].
  %%CITATION = MNRAA,291,L33;%%


%\cite{Komatsu:2010fb}
\bibitem{Komatsu:2010fb}
  E.~Komatsu {\it et al.}  [WMAP Collaboration],
  %``Seven-Year Wilkinson Microwave Anisotropy Probe (WMAP) Observations:
  %Cosmological Interpretation,''
  Astrophys.\ J.\ Suppl.\  {\bf 192}, 18 (2011)
  [arXiv:1001.4538 [astro-ph.CO]].
  %%CITATION = APJSA,192,18;%%

%\cite{Geng:2011ka}
\bibitem{Geng:2011ka}
  C.~-Q.~Geng, C.~-C.~Lee and E.~N.~Saridakis,
  %``Observational Constraints on Teleparallel Dark Energy,''
  JCAP {\bf 1201}, 002 (2012)  [arXiv:1110.0913 [astro-ph.CO]].
  %%CITATION = ARXIV:1110.0913;%%

%\cite{Bamba:2010wb}
\bibitem{Bamba:2010wb}
  K.~Bamba, C.~-Q.~Geng, C.~-C.~Lee and L.~-W.~Luo,
  %``Equation of state for dark energy in $f(T)$ gravity,''
  JCAP {\bf 1101}, 021 (2011)  [arXiv:1011.0508 [astro-ph.CO]].
  %%CITATION = ARXIV:1011.0508;%%

%\cite{Yang:2010xq}
\bibitem{Yang:2010xq}
  L.~Yang, C.~-C.~Lee, L.~-W.~Luo and C.~-Q.~Geng,
  %``Observational Constraints on Exponential Gravity,''
  Phys.\ Rev.\ D {\bf 82}, 103515 (2010)  [arXiv:1010.2058 [astro-ph.CO]].
  %%CITATION = ARXIV:1010.2058;%%


%\cite{Perivolaropoulos:2004yr}
\bibitem{Perivolaropoulos:2004yr}
  L.~Perivolaropoulos,
  %``Constraints on linear-negative potentials in quintessence and phantom
  %models from recent supernova data,''
  Phys.\ Rev.\  D {\bf 71}, 063503 (2005)
  [arXiv:astro-ph/0412308].
  %%CITATION = PHRVA,D71,063503;%%

%\cite{Li:2009bn}
\bibitem{Li:2009bn}
  M.~Li, X.~-D.~Li, S.~Wang and X.~Zhang,
  %``Holographic dark energy models: A comparison from the latest observational data,''
  JCAP {\bf 0906}, 036 (2009)  [arXiv:0904.0928 [astro-ph.CO]].
%%CITATION = ARXIV:0904.0928;%%


\end{thebibliography}
\end{document}